# Optimization of an electromagnetics code with multicore wavefront diamond blocking and multi-dimensional intra-tile parallelization

Tareq M. Malas[1]    Julian Hornich[2,4]    Georg Hager[3]    Hatem Ltaief[1]    Christoph Pflaum[2,4]    David E. Keyes[1]

[1]Extreme Computing Research Center
King Abdullah University of Science and Technology
Thuwal, Saudi Arabia
{tareq.malas,hatem.ltaief,david.keyes}@kaust.edu.sa

[2]Department of Computer Science
Friedrich-Alexander University Erlangen-Nürnberg (FAU)
Erlangen, Germany
{julian.hornich,christoph.pflaum}@fau.de

[3]Erlangen Regional Computing Center (RRZE)
Friedrich-Alexander University Erlangen-Nürnberg (FAU)
Erlangen, Germany
georg.hager@fau.de

[4]Erlangen Graduate School in Advanced Optical Technologies (SAOT),
Friedrich-Alexander University Erlangen-Nürnberg (FAU)
Erlangen, Germany

*Abstract*—Understanding and optimizing the properties of solar cells is becoming a key issue in the search for alternatives to nuclear and fossil energy sources. A theoretical analysis via numerical simulations involves solving Maxwell's Equations in discretized form and typically requires substantial computing effort. We start from a hybrid-parallel (MPI+OpenMP) production code that implements the Time Harmonic Inverse Iteration Method (THIIM) with Finite-Difference Frequency Domain (FDFD) discretization. Although this algorithm has the characteristics of a strongly bandwidth-bound stencil update scheme, it is significantly different from the popular stencil types that have been exhaustively studied in the high performance computing literature to date. We apply a recently developed stencil optimization technique, multicore wavefront diamond tiling with multi-dimensional cache block sharing, and describe in detail the peculiarities that need to be considered due to the special stencil structure. Concurrency in updating the components of the electric and magnetic fields provides an additional level of parallelism. The dependence of the cache size requirement of the optimized code on the blocking parameters is modeled accurately, and an auto-tuner searches for optimal configurations in the remaining parameter space. We were able to completely decouple the execution from the memory bandwidth bottleneck, accelerating the implementation by a factor of three to four compared to an optimal implementation with pure spatial blocking on an 18-core Intel Haswell CPU.

*Keywords*-Parallel programming, electromagnetics, performance analysis, stencils

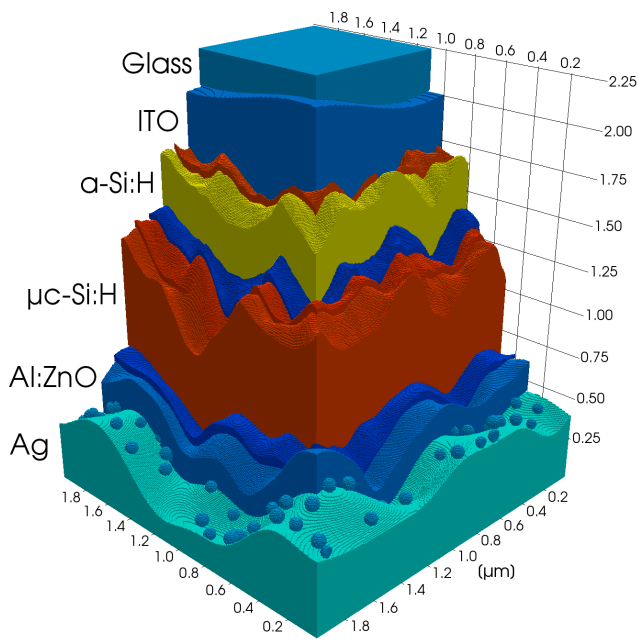

Fig. 1: Cross-section of a sample simulation setup of a tandem thin-film solar cell. The amorphous (a-Si:H) and microcrystalline silicon ($\mu$c-Si:H) layers have textured surfaces to increase the light trapping ability of the cell. $SiO_2$ nano particles are incorporated to further increase light scattering at the bottom electrode (Ag).

## I. INTRODUCTION

### A. *Photovoltaic devices and Maxwell's Equations*

Photovoltaic (PV) devices play a central role in the recent transition from nuclear and fossil fuels to more environmentally friendly sources of energy. There exist various different PV technologies, ranging from well-established polycrystalline silicon solar cells with thicknesses up to 300 $\mu$m to more recent thin-film technologies with active layer thicknesses of only 1 $\mu$m or less. To improve these thin-film PV devices and make them more competitive against other renewable energy sources, optimization of their optical properties is decisive. The importance of an optimal collection of the incident light can be seen in any of the recently developed most efficient solar cell designs [1].

To understand the effects of different light trapping techniques incorporated into PV devices and improve upon them, detailed optical simulations are necessary. The simulation code we investigate here uses the Time Harmonic Inverse Iteration Method (THIIM) [2], which is based on the staggered grid algorithm originally proposed by Yee [3] and uses the Finite-Difference Frequency Domain (FDFD) method to discretize Maxwell's Equations.

Although the algorithm performs stencil-like updates on the electric and magnetic field components, the code is different from the well-studied standard stencil benchmarks such as the 7-point stencil with constant coefficients emerging from a discretized Laplace operator. Multiple components per grid cell are involved since six coupled partial differential equations discretized by finite differences must be solved. It uses staggered grids, which results in non-symmetric data dependencies that affect the tiling structure. The loop kernels of the simulation code have very low arithmetic intensity (0.18 flops/byte) for the naive implementation), leading to a memory bandwidth-starved situation. The number of bytes per grid cell is large (40 double-complex numbers), which makes it difficult to maintain a sufficiently small working set to have the necessary in-cache data reuse for decoupling from the main memory bandwidth bottleneck.

The time-harmonic variants of Maxwell's equations are given by

$$i\omega\hat{\mathbf{E}} = \frac{1}{\varepsilon}\nabla \times \hat{\mathbf{H}} - \frac{\sigma}{\varepsilon}\hat{\mathbf{E}} \ , \quad (1)$$

$$i\omega\hat{\mathbf{H}} = -\frac{1}{\mu}\nabla \times \hat{\mathbf{E}} - \frac{\sigma^\star}{\mu}\hat{\mathbf{H}} \ , \quad (2)$$

with permittivity $\varepsilon$, permeability $\mu$, the electric and magnetic conductivities $\sigma$ and $\sigma^\star$, and the frequency of the incident plane wave $\omega$. The time-independent electric and magnetic field components are related to the time-dependent fields by $\vec{\mathbf{E}} = \hat{\mathbf{E}}e^{i\omega\tau}$ and $\vec{\mathbf{H}} = \hat{\mathbf{H}}e^{i\omega\tau}$.

After discretization of Maxwell's equations in time and space the following iterative scheme is obtained:

$$\frac{e^{i\omega\tau}\hat{\mathbf{E}}_h^{n+1} - \hat{\mathbf{E}}_h^n}{\tau} = \frac{1}{\varepsilon}\nabla_h \times \hat{\mathbf{H}}_h^{n+\frac{1}{2}} e^{i\omega\frac{\tau}{2}} - \frac{\sigma}{\varepsilon}\hat{\mathbf{E}}_h^{n+1} e^{i\omega\tau} + \mathbf{S}_E \ , \quad (3)$$

$$\frac{e^{i\omega\frac{\tau}{2}}\hat{\mathbf{H}}_h^{n+\frac{1}{2}} - e^{-i\omega\frac{\tau}{2}}\hat{\mathbf{H}}_h^{n-\frac{1}{2}}}{\tau} = -\frac{1}{\mu}\nabla_h \times \hat{\mathbf{E}}_h^n - \frac{\sigma^\star}{\mu}\hat{\mathbf{H}}_h^{n+\frac{1}{2}} + \mathbf{S}_H \ , \quad (4)$$

with time step $\tau$, time step index $n$ and source terms $\mathbf{S}_E$ and $\mathbf{S}_H$. To model materials with negative permittivity ($\varepsilon < 0$, e.g., silver electrodes) the THIIM method applies a "back iteration" scheme to the electric field components of the corresponding grid points:

$$\frac{e^{i\omega\tau}\hat{\mathbf{E}}_h^n - \hat{\mathbf{E}}_h^{n+1}}{\tau} = \frac{1}{\varepsilon}\nabla_h \times \hat{\mathbf{H}}_h^{n+\frac{1}{2}} e^{i\omega\frac{\tau}{2}} - \frac{\sigma}{\varepsilon}\hat{\mathbf{E}}_h^{n+1} + \mathbf{S}_E \ . \quad (5)$$

With this method, the optical constants of any material can be used directly in the frequency domain without the need for any approximation or auxiliary differential equations [4], [5], [6], [7]. THIIM has proven to be numerically stable and give accurate solutions for setups with metallic back contacts [8], [9] and also for the simulation of plasmonic effects, e.g. around silver nano wires [10].

A perfectly matched layer (PML) is used to allow absorption of outgoing waves, employing the split-field technique originally presented by Berenger [11]: All six $\hat{\mathbf{E}}$ and $\hat{\mathbf{H}}$ field components are split into two parts each. For example, the $\hat{\mathbf{E}}_\mathbf{x}$ component of equation 1 is split into $\hat{\mathbf{E}}_\mathbf{x} = \hat{\mathbf{E}}_\mathbf{xy} + \hat{\mathbf{E}}_\mathbf{xz}$, resulting in two equations:

$$(i\omega\varepsilon + \sigma_y)\hat{\mathbf{E}}_\mathbf{xy} = \frac{\partial}{\partial y}\left(\hat{\mathbf{H}}_\mathbf{zx} + \hat{\mathbf{H}}_\mathbf{zy}\right), \quad (6)$$

$$(i\omega\varepsilon + \sigma_z)\hat{\mathbf{E}}_\mathbf{xz} = -\frac{\partial}{\partial z}\left(\hat{\mathbf{H}}_\mathbf{yx} + \hat{\mathbf{H}}_\mathbf{yz}\right). \quad (7)$$

For all six vector components this procedure is performed on Equations 3, 4 and 5 resulting in a total of 12 coupled equations.

In order to overcome the problem of the representation of complicated light-trapping geometries, such as rough interfaces between layers or curved particle surfaces, the Finite Integration Technique (FIT) [12] is applied on the rectangular structured grids. FIT allows to accurately treat curved interfaces by integrating the material data on an unstructured tetrahedron grid and mapping the data back to the structured grid.

Figure 1 shows a sample simulation setup of a thin-film tandem solar cell that can be simulated by the methods mentioned above [13]. The amorphous and microcrystalline silicon layers are used to absorb different ranges of the incident spectrum. Their surfaces are etched to increase the trapping of light inside the cell. Atomic force microscopy is used to obtain height information that is then introduced between the layers in the simulation. Additionally, at the back electrode (Ag) $SiO_2$ nano particles can be deposited to increase the scattering of light. For such a setup PML boundary conditions are chosen vertically. Horizontally periodic boundary conditions are used.

*B. Contribution*

This work makes the following contributions: We optimize the multi-threaded (OpenMP-parallel) part of the THIIM code using temporal blocking. Our multi-dimensional intra-tile parallelization implementation shows a significant reduction in the cache block size requirement, providing sufficient data reuse in the cache to decouple from the main memory bandwidth bottleneck. As a result, we obtain a $3\times$–$4\times$ speedup compared to an efficient spatially blocked code. In addition to the performance improvements, our results show significant memory bandwidth savings of 38%–80% off the available memory bandwidth, making it immune to more memory bandwidth-starved systems. Via appropriate cache block size and code balance models we prove that cache block sharing is essential for decoupling from the memory bandwidth bottleneck. We validate these models by analyzing different tile sizes and measuring relevant hardware performance counters.

In the current usage scenarios for this code, the overhead caused by MPI communication is negligible or can be hidden by a dedicated communication thread. An in-depth analysis of the performance implications of communication overhead for the temporally blocked variant is out of scope for this paper and left for future work.

## II. APPROACH: MULTI-DIMENSIONAL CACHE BLOCK SHARING

### A. Background

Temporal blocking is a well-known technique to reduce the data traffic to the main memory for memory bandwidth-starved stencil computations. It allows the code to iterate multiple times over a subdomain that fits into a cache. In recent years, diamond tiling has moved into the focus of research on temporal blocking [14], [15], [16], [17]. Diamond tiling provides a convenient and unified data structure to maximize the in-cache data reuse [14], has low synchronization requirements, allows more concurrency in tile updates, and can be utilized to perform domain decomposition in a distributed memory setup [17].

Wavefront blocking is another important temporal blocking technique, introduced by Lamport [18]. It maximizes the data reuse in a given space-time block as long as the wavefront tile fits in the desired cache memory level.

The combination of diamond tiling with wavefront bocking in three-dimensional problems is becoming more popular. These techniques have recently been shown to yield good performance [16], [17], and diamond tiling has been implemented in the PLUTO framework [19]. Wavefront diamond blocking is usually applied to the outer two space dimensions. The fast moving (inner) dimension is usually left untouched for better performance, as confirmed in [16], [19], [20]. The contiguous memory access of a long inner loop is important for efficient hardware data prefetching and better utilization of the CPU execution pipelines.

Multicore-aware cache block sharing techniques, introduced in [21], are another way to reduce the data traffic in bandwidth-starved situations. Cache block sharing among the threads of the processor reduces the number of tiles required to fit in the cache memory. As a result, larger tiles can fit in the cache memory to provide more in-cache data reuse. This technique is particularly important for the THIIM stencil in this paper. The THIIM stencil requires many bytes per grid cell, which makes it challenging to fit sufficiently large tiles in the cache memory. We have introduced a more advanced cache block sharing technique in [22], where we propose multi-dimensional intra-tile parallelization to achieve a further reduction in the tile size requirements and maintain architecture-friendly memory access patterns.

Our experiments require a system that allows full control over the tunable parameters of a temporally blocked stencil algorithm. The open source system provided by Malas *et al.* [17], [22], called *Girih*, provides these options. Girih uses wavefront-diamond tiling with multi-dimensional intra-tile parallelization to construct a Multi-threaded Wavefront Diamond blocking (MWD) approach. It allows running multiple threads per cache block, fitting larger blocks in the cache memory to reduce the data traffic to main memory. The implementation leverages the performance counter API of the LIKWID multicore tools collection [23] to measure the data traffic in the memory hierarchy via the provided hardware counters.

Threads are assigned to the tiles in Thread Groups (TGs), similar to [21]. Multiple TGs can run concurrently, updating different tiles and observing inter-tile dependencies. The TG size parameter provides a controllable tradeoff between concurrency and sharing of the loaded data from memory among the available threads.

Diamond tiles are dynamically scheduled to the available TGs. A First In First Out (FIFO) queue keeps track of the available diamond tiles for updating. TGs pop tiles from this queue to update them. When a TG completes a tile update, it pushes to the queue its dependent diamond tile, if that has no other dependencies. The queue update is performed in an OpenMP critical region to avoid race conditions. Since the queue updates are performed infrequently, the lock overhead is negligible.

We use the auto-tuner in the Girih system to select the diamond tile size, the wavefront tile width, and the TG size in all dimensions to achieve the best performance. To shorten the auto-tuning process, the parameter search space is narrowed down to diamond tiles that fit within a predefined cache size range using a cache block size model.

### B. Multi-dimensional intra-tile parallelization

Figure 2 shows the diamond tiling implementation of the THIIM stencil kernel. We split the $\hat{\mathbf{H}}$ and $\hat{\mathbf{E}}$ field updates in the figure as they have different data dependency directions. The $\hat{\mathbf{H}}$ and $\hat{\mathbf{E}}$ fields have dependencies over the positive and negative directions, respectively, as illustrated in Fig. 3. Splitting the fields allows more data reuse in the diamond tile and provides proper tessellation of diamond tiles. As a result, a full diamond tile update starts and ends with an $\hat{\mathbf{E}}$ field update. The horizontal (blue) lines in Fig. 3 divide the components in three regions, which can be handled by three threads. See below for details.

The extruded diamond tile is shown in Figure 4. We perform the wavefront traversal along the *z* dimension (outer dimension) and the diamond tiling along the *y* dimension (middle dimension). We do not tile the *x* dimension (fast moving dimension), as we split its work among multiple threads with simultaneous updates in the TG.

The staggered grid and multi-component nature of this application requires different intra-tile parallelization strategies than "standard" structured grid implementations. We use a Fixed-Execution to Data (FED) wavefront parallelization approach [22], which always assigns the same grid points to each thread while the wavefront traverses the tile. This idea maximizes the data reuse in the thread-private caches, since only boundary data instead of the full tile has to travel between threads. The corresponding performance improvement is very limited for simple stencils, but the THIIM stencil and high-order stencils do benefit from it.

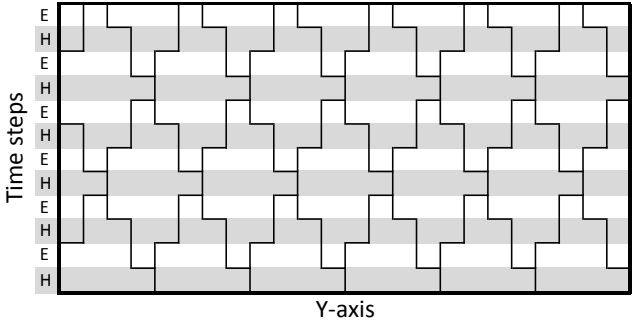

Fig. 2: Diamond tile shape along the *y* dimension for the THIIM stencil. Although the $\hat{\mathbf{H}}$ and $\hat{\mathbf{E}}$ fields are updated in the same iteration of the simulation code, we split them in our tiling implementation to achieve better data reuse and better diamond tile tessellation.

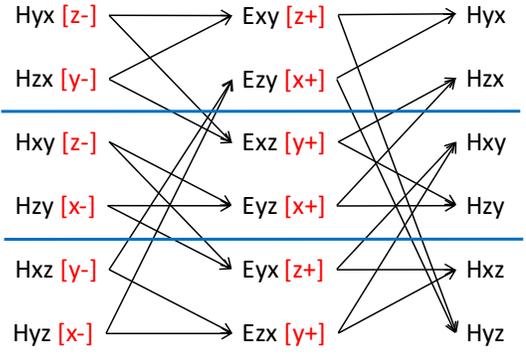

Fig. 3: $\hat{\mathbf{H}}$ and $\hat{\mathbf{E}}$ field dependencies of the THIIM stencil kernel. Each field is updated by reading six domain-sized arrays of the other field. The arrows indicate dependencies over the same location in the grid and a unit index offset. The (red) labels in square brackets indicate the axis and the direction of the offset. The (blue) horizontal lines split the components in three regions to indicate the components update parallelism using three threads.

We allow a concurrent update of the *x* dimension grid cells by the threads in the TG while the data is in cache. This handling of the *x* dimension has two advantages: it reduces the pressure on the private caches of the threads, and it maintains data access patterns that allow for efficient use of hardware prefetching and the Translation Lookaside Buffer (TLB).

The fixed amount of work per time step in the *z* and *x* dimensions leads to good load balancing, but parallelizing the diamond tile along the *y* dimension can be inefficient since the odd number of grid points at every other time step in the diamond tile makes load balancing impossible for more than one thread along the *y* dimension. Doubling the diamond tile width is possible, but it would result in doubling the cache block size without increasing the data reuse. Moreover, a load-balanced implementation cannot make the intra-tile split parallel to the time dimension, so more data will have to move between the private caches of the threads. As a result, we do not perform intra-tile parallelization along the diamond tiling dimension for this stencil.

We exploit the concurrency in the field component updates by adding a further dimension of thread parallelism. Each field update can update six fields concurrently. We parameterize our code to allow 1, 2, 3, and 6-way parallelism in the field update so that the auto-tuner selects the most performance-efficient configuration. For example, Fig. 3 shows a case of parallelizing the components update using three threads.

In our spatial blocking and MWD benchmark implementations we use homogeneous Dirichlet boundary conditions in all dimensions to study the performance improvements of our techniques. We expect no significant changes in performance with periodic boundary conditions.

### III. DETAILED ANALYSIS OF THE STENCIL CODES

Here we analyze the data traffic requirements per lattice site update (i.e., the code balance) of the stencil code for the naïve, spatially blocked, and temporally blocked variants.

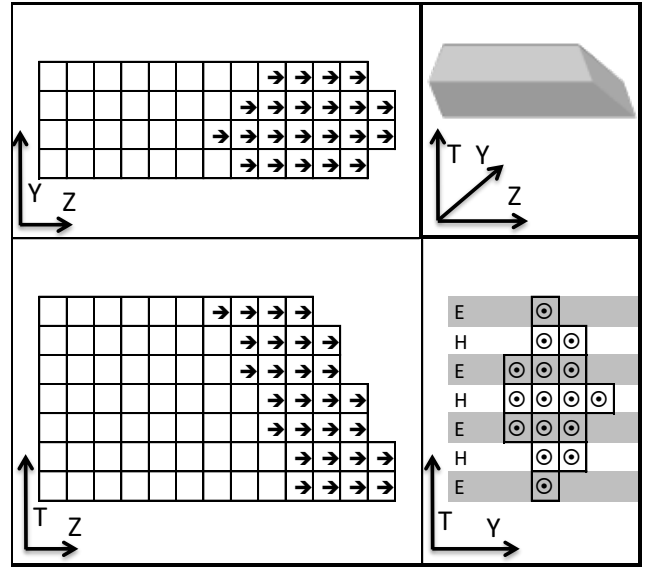

Fig. 4: Extruded diamond tiling of the THIIM kernels, showing an example of $D_w = 4$ and $W_w = 4$. The data dependencies of the $\hat{\mathbf{H}}$ and $\hat{\mathbf{E}}$ fields allow more data reuse in the wavefront.

As described above, six components each are used for the electric field $\hat{\mathbf{E}}$ and the magnetic field $\hat{\mathbf{H}}$. We show the code of two component updates in the THIIM stencil in Listings 1 and 2. The remaining three and seven components updates have very similar memory access and computation patterns. The $H_{XY}$ update in Listing 1 uses three coefficient arrays (tHyx, cHyx, SrcHy) and the $H_{ZX}$ update in Listing 2 uses two coefficient arrays (tHzx, cHzx). Overall (i.e., considering all component updates) this results in $4 \cdot 3 + 8 \cdot 2 = 28$ domain-sized arrays for the coefficients. In total, $12 + 28 = 40$ domain-sized arrays have to be stored using double-complex numbers, leading to a storage requirement per grid cell of $16 \cdot 40 \text{ bytes} =$

Listing 1: Kernel for the magnetic field $H_{YX}$ component update. Accesses with index shifts along the outer dimension are highlighted. Similar computations and memory access patterns are performed in $H_{XY}$, $E_{YX}$, and $E_{XY}$ updates.

```
for(k=zb; k<ze; k++) {
 for(j=yb; j<ye; j++) {
  ib=2*((k*Ny+j)*Nx+xb); ie=2*((k*Ny+j)*Nx+xe);
  for(i=ib; i<ie; i+=2) {
   ishift = i+2*(-Nx*Ny);
   Re=Exy[i]-Exy[ishift]+Exz[i]-Exz[ishift];
   Im=Exy[i+1]-Exy[ishift+1]+Exz[i+1]-Exz[ishift+1];
   t=Hyx[i]*tHyx[i]-Hyx[i+1]*tHyx[i+1]+SrcHy[i]
       -cHyx[i]*Re+cHyx[i+1]*Im;
   Hyx[i+1]=Hyx[i]*tHyx[i+1]+Hyx[i+1]*tHyx[i]
       +SrcHy[i+1]-cHyx[i]*Im-cHyx[i+1]*Re;
   Hyx[i] = t; }}}
```

Listing 2: Kernel for the magnetic field $H_{ZX}$ component update. Accesses with index shifts along the middle dimension are highlighted. Similar computations and memory access patterns are performed in $H_{ZY}$, $H_{XZ}$, $H_{YZ}$, $E_{XZ}$, $E_{YZ}$, $E_{ZX}$, $E_{ZY}$ updates. The offset direction in ishift variable differs in the components updates as presented in Fig. 3.

```
for(k=zb; k<ze; k++) {
 for(j=yb; j<ye; j++) {
  ib=2*((k*Ny+j)*Nx+xb); ie=2*((k*Ny+j)*Nx+xe);
  for(i=ib; i<ie; i+=2) {
   ishift = i+2*(-Nx);
   Re=Exy[ishift]-Exy[i]+Exz[ishift]-Exz[i];
   Im=Exy[ishift+1]-Exy[i+1]+Exz[ishift+1]-Exz[i+1];
   t = Hzx[i]*tHzx[i]-Hzx[i+1]*tHzx[i+1]
       -cHzx[i]*Re+cHzx[i+1]*Im;
   Hzx[i+1] = Hzx[i]*tHzx[i+1]+Hzx[i+1]*tHzx[i]
       -cHzx[i]*Im-cHzx[i+1]*Re;
   Hzx[i] = t; }}}
```

640 bytes per grid cell.

### A. Naïve kernel arithmetic intensity

We count the total floating-point operations per Lattice-site Update (LUP) in the stencil code. The loop nests in Listings 1 and 2 perform 22 flops and 20 flops, respectively. In total we count $4 \cdot 22 + 8 \cdot 20 = 248$ Double Precision (DP) flops/LUP. For calculating the data traffic we note that the loop in Listing 1 writes two double precision numbers, reads twelve numbers with no index shift, and reads four numbers with an outer dimension index shift (ishift). If we assume that all accesses to arrays with an outer dimension index shift (in Listing 1 these are Exy and Exz) actually go to main memory we have a total traffic of 18 double precision numbers in this loop. Whether this is true or not depends on the problem size: if two successive *x-y* layers of those grids fit into the cache, the shifted and non-shifted accesses to the same arrays come at half the data transfer cost because the access with the smaller index comes from cache. This reasoning is well known in stencil optimizations [24], [25]. At a problem size of $512^3$ two layers take up $512^2 \cdot 16 \cdot 2 = 8$ MiB of cache *per thread and per array*, which exceeds the available cache size by far. See the next section on how this can be corrected.

The code in Listing 2 writes two numbers and reads ten numbers without large index shifts. The shifted accesses to Exz and Exy can be ignored since the shift is only along the middle dimension, and two rows of the data easily fit into some cache. The third variant of array updates is identical to the second in terms of data transfers since it has a very small shift of $-2$ along the inner dimension only.

Overall we thus have a code balance of

$$B_C = 4 \cdot (18 + 12 + 12) \cdot 8 \text{ bytes/LUP} = 1344 \text{ bytes/LUP}, \quad (8)$$

leading to an arithmetic intensity of $I = 248/1344$ flops/byte $= 0.18$ flops/byte, which results in very high pressure on the main memory bandwidth.

### B. Spatial blocking arithmetic intensity

The total load/store operations to memory can be reduced by standard spatial blocking techniques, which establish "layer conditions" along the outer grid dimensions (see, e.g., [25] and references therein). Spatial blocking results in a reduction of the memory traffic in the four loop nests that are structured as shown in Listing 1 by four double precision numbers each, if the blocking sizes in the inner and/or middle dimensions are chosen such that two successive layers of an array with index shifts in the outer dimension (highlighted in the listing) fit into a cache. The new code balance is thus

$$B_C = 4 \cdot ([18-4] + 12 + 12) \cdot 8 \text{ bytes/LUP} = 1216 \text{ bytes/LUP}, \quad (9)$$

and the arithmetic intensity becomes $I = 248/1216$ flops/byte $= 0.20$ flops/byte. The spatial blocking optimization improves the performance of the code by a mere 10% because the main contributors to the data traffic are not the electric and magnetic fields but the coefficient arrays. Spatial blocking is not effective for these because they are accessed with no temporal locality.

We can now predict the maximum performance for optimal spatial blocking using a simple bottleneck model [26]: The limit due to the maximum memory bandwidth $b_S$ of the CPU is $P_{\text{mem}} = b_S/B_C$. The Haswell chip we used for our experiments has $b_S \approx 50$ GB/s (see Sect. IV-A), hence

$$P_{\text{mem}} = \frac{b_S}{B_C} = \frac{50 \text{ GB/s}}{1216 \text{ bytes/LUP}} = 41 \text{ MLUP/s}. \quad (10)$$

This prediction is in very good agreement with the measurements. See Sect. IV for details.

### C. Diamond tiling arithmetic intensity and cache size requirements

The performance of temporal blocking techniques relies on a reduction of data traffic, especially to and from main memory. Data traffic models are very useful for understanding the expected or observed performance gains. We build a cache block size model and a code balance model based on [22]. The cache block size model estimates the maximum tile size that fits in the cache memory for this application by counting the working data set in the diamond-wavefront tile, such as the one shown in the y-z plane in Fig. 4. The total required number of bytes per tile is:

$$C_s = 16 \cdot N_x \cdot \left[ 40 \cdot \left( \frac{D_w^2}{2} + D_w \cdot (B_Z - 1) \right) + 12 \cdot (D_w + W_w) \right]. \quad (11)$$

Each point in the diamond-wavefront tile extends over the full length of the x dimension ($N_x$) with double-complex values (8·2 bytes). The area of the wavefront-tile is $\frac{D_w^2}{2} + D_w \cdot (B_Z - 1)$, which depends on the diamond tile width ($D_w$) and the tile size along the z dimension ($B_Z$). Since each grid cell requires loading 12 components and 32 coefficients, we multiply the wavefront-diamond tile area by 40 numbers per grid cell. Finally, the $12 \cdot (D_w + W_w)$ part corresponds to the neighbor access of the 12 components around the wavefront-diamond tile, where the wavefront tile width $W_w = D_w + B_Z - 1$. For example, in Fig. 4 we have $D_w = 4$, $B_Z = 4$, and $W_w = 7$, so we have $C_s = 14912 \cdot N_x$ bytes per cache block.

For the code balance model we have to estimate the potential reduction of memory bandwidth pressure by temporal blocking. If the tile fits entirely in the L3 cache, the code loads each grid point once from main memory and stores it back only after completing the wavefront updates. We count the total reads and writes per diamond tile and divide by the diamond area (i.e., data reuse). Each diamond update consists of writing six $\hat{\mathbf{H}}$ field components per cell at full diamond width ($D_w$) and writing six $\hat{\mathbf{E}}$ field components per cell at $D_w - 1$. In total, each diamond requires $6 \cdot (2 \cdot D_w - 1)$ writes. The diamond tile requires reading 40 numbers per cell and accessing the neighbors of the 12 components ($40 \cdot D_w + 12$). The diamond area is ($D_w^2/2$). The code balance for double-complex numbers of the kernel is thus:

$$B_C = \frac{16 \cdot [6 \cdot (2 \cdot D_w - 1) + (40 \cdot D_w + 12)]}{D_w^2/2} \frac{\text{bytes}}{\text{LUP}}. \quad (12)$$

We validate our models and study the potential impact of our temporal blocking techniques using our Single-threaded Wavefront Diamond blocking (1WD) implementation. Figure 5 shows the model predictions (solid black lines) of the code balance and cache block size and the code balance measurements (dashed blue lines). The latter is based on a direct measurement of the memory data traffic via hardware performance counters. We test four diamond tile widths (4, 8, 12, and 16). The red vertical lines indicate the estimated usable block size in the L3 cache of the Haswell processor (as a rule of thumb we assume that half the overall cache size, i.e., 22.5 MiB, is available for tile data). Figures 5a–5c correspond to three wavefront width sizes ($B_Z = 1$, 6, and 9), where more concurrency is achievable along the z dimension at the cost of using a larger cache block size. We perform our tests at a grid of size $480^3$ in the 18-core Haswell processor using a single core and a single cache block.

The measurements show that the model accurately predicts the usable cache block size. The measured code balance diverges from the model when more than half of the L3 cache is used (the right side of the vertical red line), which is expected.

Our results also emphasize the importance of multi-dimensional intra-tile parallelism compared to parallelizing the wavefront only, where the maximum number of threads per tile is restricted by the wavefront tile width. Using $B_Z = 6$ would require three thread groups at the Haswell processor. As a result, the minimum diamond width $D_w = 4$ requires a cache block size $C_s = 30$ MiB, which exceeds the available cache memory. Although the cache block size of $B_Z = 9$ would fit in the L3 cache at $D_w = 4$, it cannot use larger diamond tiles, to enable more data reuse. On the other hand, our approach provides parallelism along the other dimensions without increasing the cache block size (i.e., it uses smaller wavefront tile widths), which saves space for larger diamond tiles. For example, we can set $B_Z = 1$ and use nine threads per cache block along the other dimensions. This setup provides a $D_w = 8$ that uses $C_s = 20$ MiB, allowing more data reuse within the usable cache block size limit.

## IV. RESULTS

We present performance results for the spatially blocked code, 1WD, and MWD with full parameters auto-tuning to show the performance improvements. In order to get more insight into performance properties we show thread scaling at fixed grid size and full socket performance at increasing grid size (cubic domain). Since 1WD generally performs better than PLUTO and Pochoir as we show in [22], we do not implement the THIIM stencil in their frameworks to compare their performance. We also present results using different thread group sizes to show the impact of the cache block sharing over the memory transfer volume and the memory bandwidth.

### A. Test hardware

We conducted all experiments on an Intel Haswell EP CPU (18-core Xeon E5-2699 v3, 2.3 GHz, 45 MB L3 cache, and 50 GB/s of applicable memory bandwidth), which features the most recent microarchitecture available to date in an Intel Xeon processor. The "Turbo Mode" feature was disabled, i.e., the CPUs ran at their nominal clock speed of 2.3 GHz, to avoid performance fluctuations. The chip was set up in a standard configuration with "Cluster on Die" (CoD) mode disabled, meaning that the full chip was a single ccNUMA memory domain with 18 cores and a shared L3 cache. The documented clock slowdown feature with highly optimized AVX code [27] was not observed on this machine with any of our codes. Simultaneous multi-threading (SMT) was not used.

### B. Thread scaling results

We present performance results of the THIIM kernel at increasing number of threads for a fixed problem size in Figure 6a. We also show the memory bandwidth measurements in Fig. 6b, measured code balance in Fig. 6c, and the auto-tuned MWD diamond width parameter in Fig. 6d.

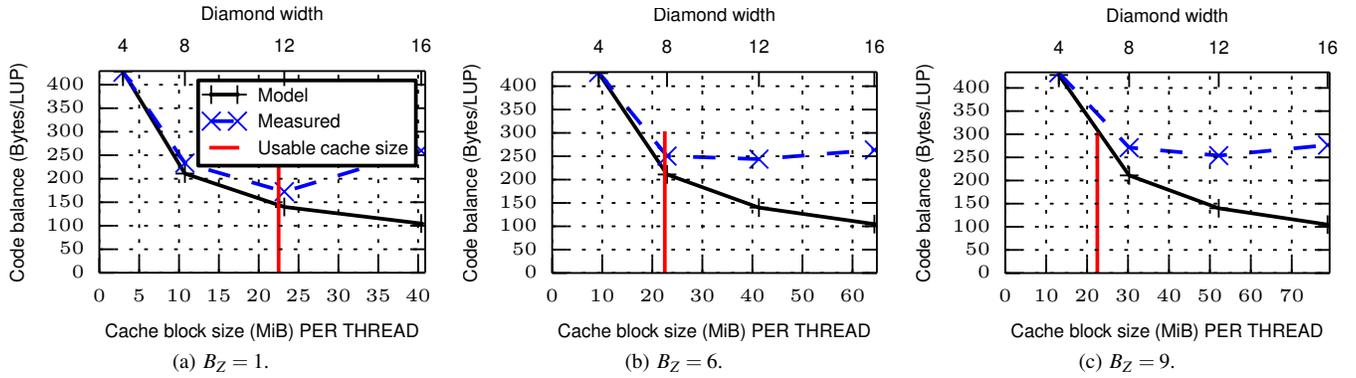

Fig. 5: The cache block size requirements of the application's kernels at three wavefront widths ($B_Z$). We use an 18-core Haswell at grid size $480^3$, running a single thread with the 1WD approach. Smaller wavefront tile widths, which provide less concurrency along the $z$ dimension, enable more data reuse.

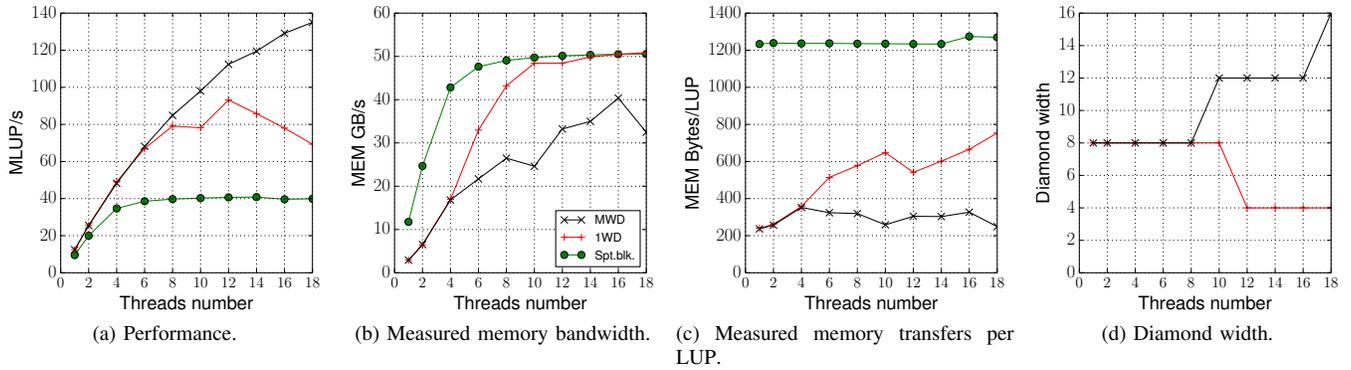

Fig. 6: The THIIM stencil performance and memory transfer measurements, comparing 1WD, MWD, and spatially blocked code variants on an 18-core Haswell socket at increasing number of threads using grid size $384^3$.

The spatially blocked code saturates the memory interface already with six cores, resulting in a performance of around 40 MLUP/s. This is in very good agreement with the bandwidth-based prediction of 41 MLUP/s that was derived in Sect. III-B. Using separate cache blocks per thread in 1WD alleviates the memory bandwidth pressure and achieves better performance than spatial blocking code at smaller thread counts, but the cache is too small to accommodate sufficient blocks at larger thread counts so that a performance drop is observed beyond twelve cores. This can be seen more clearly in Fig. 6b: 1WD goes into bandwidth saturation at ten cores. In contrast, MWD does not saturate the memory bandwidth and can still profit from more cores up to the chip limit, showing a parallel efficiency of about 75% on the full chip. It can maintain a low code balance of 200–400 bytes/LUP for all thread counts (see Fig. 6c). The comparison of diamond width parameters selected by the autotuner in Fig. 6d is quite revealing: at larger core counts, 1WD requires smaller diamonds to meet the stringent cache size limit per core, whereas MWD can employ larger diamonds due to several threads sharing a diamond tile for wavefront updates.

### C. Increasing grid size results

Although thread scaling, as shown in the previous section, reveals many interesting features of the 1WD and MWD algorithms, it is also instructive to study their behavior with changing problem size. We therefore present performance results of the THIIM kernel at different (cubic) grid sizes in Figure 7a, ranging from 64 to 512 with an increment of 64. We also show the auto-tuned MWD intra-tile parallelization parameters in Fig. 7b, the memory bandwidth measurements in Fig. 7c, and measured code balance in Fig. 7d.

1WD performance decays at larger grid size because of the increasing cache requirements as the leading dimension grows. The rise in the memory transfer volume seen in Fig. 7d suggests that the larger cache blocks cause more capacity misses in the L3 cache. Our auto-tuner selects a very small $D_W = 4$ at all grid sizes of 1WD, which already exceeds the available cache memory.

Our MWD implementation is decoupled from the memory bandwidth bottleneck over the full range of problem sizes. Compared to the spatially blocked code it has a 6× lower code balance, resulting in a 3×–4× speedup. The memory bandwidth measurements in Fig. 7c show that our approach is immune to even more memory bandwidth-starved situations, where the machine balance (ratio of memory bandwidth to computational performance) would be lower.

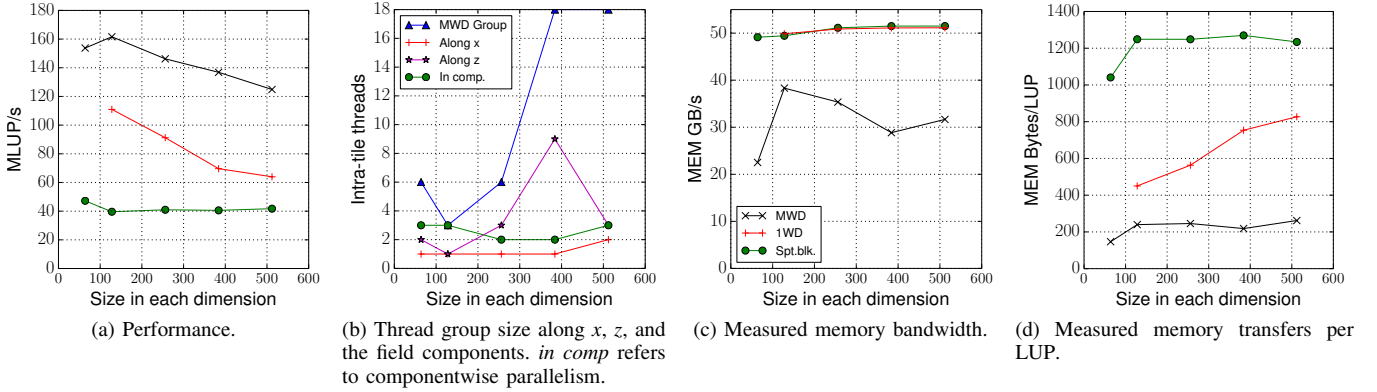

Fig. 7: The THIIM stencil kernel performance and memory transfer measurements, comparing 1WD, MWD, and spatially blocked code variants on an 18-core Haswell socket at increasing cubic grid size.

The auto-tuner selects larger thread groups as the grid size increases, as shown in Fig. 7b, to reduce the cache size requirements. This allows diamond widths in the range 8–16. For all grid sizes, two or three threads are used for the parallel components update. The components parallelism is a major contributor in reducing the cache block size requirements while maintaining high intra-tile concurrency. On the other hand, parallelizing the wavefront dimension alone would result in a larger cache block size, as described in Sect. III-C.

*D. Thread group size impact on performance and memory transfers*

In this section we show the impact of the thread group size (i.e., cache block sharing) on the THIIM kernel performance in Figure 8a, on the memory bandwidth measurements in Fig. 8c, and on the code balance in Fig. 8d. We also show the tuned MWD diamond tile width in Fig. 8b.

The cases 6WD, 9WD, and 18WD are able to decouple from the memory bandwidth bottleneck at large grid sizes, allowing them to achieve similar performance. The small performance variations make the auto-tuner select different thread group sizes, as shown in the case of MWD performance at grid size of 512 in Figs. 7 and 8.

Larger thread group sizes reduce the need for cache size. As a result, increasing the thread group size allows the auto-tuner to select a larger diamond tile width, resulting in more in-cache data reuse, less memory bandwidth, and less memory transfer volumes. The 18WD version uses at least $D_W = 16$ at all grid sizes, as shown in Fig. 8b. The massive in-cache data reuse of 18WD results in saving more than 38% of the memory bandwidth at all grid sizes. On a CPU with smaller machine balance we expect an even more pronounced advantage of large thread group sizes.

## V. RELATED WORK

Stencil computations are important kernels in many scientific applications, as they appear in many Partial Differential Equation (PDE) codes. As a result, optimization techniques for stencil computations are studied extensively in the literature. Datta [28] showed results for many optimization techniques over several processors and stencil computations.

Temporal blocking techniques for stencil algorithms have been the subject of intense study over the last two decades. Applying all those approaches to an application code in order to find out the highest performing candidate is out of the question, and it is also not necessary: We provide a comprehensive comparison of MWD and its variations (FED and multi-dimensional intra-tile parallelism) with the widely accepted state-of-the-art frameworks and techniques CATS2/1WD [16], PLUTO [19], and Pochoir [20] in a companion paper [22]. We show there that MWD outperforms the other techniques significantly for four carefully chosen "corner case" stencil schemes. Thus we have restricted ourselves to 1WD and MWD for optimizing the electromagnetics code.

Most of the proposed temporal blocking algorithms use separate cache block per thread [16], [19], [20], [29]. Our work shows the inefficiency of these techniques for memory-starved stencils in contemporary CPUs. On the other hand, cache block sharing technologies (introduced by Wellein *et al.* [21]), achieve better performance by utilizing the shared hardware caches of modern CPUs. Recently, Shrestha *et al.* [30] introduced cache block sharing techniques within PLUTO framework to perform source-to-source transformation of the stencil codes. To the extent of our knowledge, all proposed cache block sharing temporal blocking techniques compromise tile size for intra-tile concurrency, which we show to be sub-optimal in this work.

For the simulation of solar cells the THIIM algorithm was chosen because it is a stable method for negative permittivity materials such as silver. It allows to directly use the refractive index material data in the simulation without any further approximation or an auxiliar differential equation (ADE). To model lossy dispersive media, alternative methods introduce a convolution integral between the electric flux density and the electric field. This integral is then approximated by recursive convolution [4] or piecewise linear recursive convolution [5]. Another approach expresses the convolution integral by an ADE, which is then discretized by finite differences [6]. A Z-transform (ZT) is applied to obtain field update equations [7]. The ADE and ZT approaches need reformulation for different dispersive media. All those alternative methods either introduce further unknowns or additional equations that need to be solved, increasing the memory requirement and computation time.

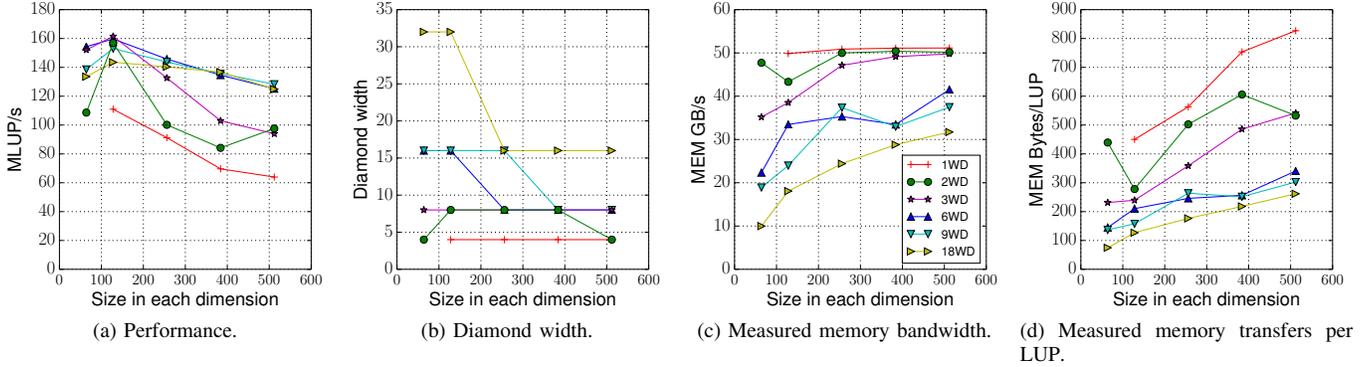

Fig. 8: The THIIM stencil kernel performance and memory transfer measurements, comparing various thread group sizes in MWD on an 18-core Haswell socket at increasing cubic grid size. The results show the ability of our approach to reduce significantly the required memory bandwidth and transfer volumes.

## VI. Conclusion and Outlook

We have applied multicore wavefront diamond temporal blocking with multi-dimensional intra-tile parallelization to a Maxwell's Equations solver used in a solar cell simulation application, achieving a 3×–4× speedup and a 38%–80% memory bandwidth saving. This stencil code has very low arithmetic intensity (0.20 flop/byte for optimal spatial blocking) and requires many bytes of storage per grid cell (640 bytes). Applying thread parallelism inside shared cache blocks as well as across electric and magnetic field components was decisive in lowering the severe cache size constraints of the code. Using a validated cache block size and code balance model we were able to describe the impact of the tiling parameters and the cache size on the memory traffic and thus limit the effort of the auto-tuner. To our knowledge, none of the existing temporal blocking techniques in the literature can achieve a similarly efficient memory bandwidth reduction for such a memory-starved stencil kernel.

The design and optimization process of solar cells requires thousands of parallel runs of this code. In order to cover the whole visible wavelength spectrum for only a single solar cell configuration, about 80–160 simulations are needed. Our performance improvements reduce the turnaround time of each individual run and also the overall cost of the computations. We believe that our approach is applicable to many algorithms with similar characteristics, i.e., where the code has significant demand for memory bandwidth and cache size.

The solar cell simulation application in this paper uses a cubic domain shape. In many applications, from climate models to reservoir models, etc., one dimension is significantly smaller than the other two, i.e., the domain is "thin." Our approach can benefit such applications significantly: Mapping the thin dimension to the leading array dimension helps both, tiling in shared memory and domain decomposition in distributed memory setups. For shared memory, we show in Eq. 11 that the cache block size is proportional to the leading dimension size, so we can use larger blocks in time with more data reuse. Although tiling a long leading dimension can also reduce the cache block size, it increases the pressure on the TLB and may lead to inefficient hardware data prefetching [25]. In distributed memory, decomposing the leading dimension is usually the most expensive, as the halo layer is not contiguous in memory. Thin domains reduce the requirement of decomposing the leading dimension while maintaining a favorable surface-to-volume ratio per subdomain. It is worth mentioning that very short leading dimensions (i.e., thin domains with less than about 50 cells) are inefficient because of bad pipeline utilization. This effect is amplified by long SIMD units, which lead to even shorter loop lengths and slow (scalar) remainder loops. In this situation the thin domain should be mapped to the middle or outer dimensions.

Our MWD work has eliminated the memory bandwidth bottleneck of this code. We are currently in the process of implementing periodic boundary conditions. These may be introduced along the *x* dimension by peeling the first and last iteration off the the *x* loop to explicitly specify the contributing grid points at the other end of the domain. Furthermore we can use the leftmost half-diamond in Fig. 2 to complete the rightmost half-diamond via a memory copy operation, and vice versa.

In the future we plan to investigate further the performance limitations within the core (in particular the SIMD vectorization) and the cache hierarchy, since the code runs at only about

5% of the theoretical peak performance of the CPU despite being cache bound. Hardware performance counter measurements and subsequent chip-level performance modeling will provide more insight here. The temporal blocking optimization will change the communication versus computation characteristics of the code, which also deserve an in-depth analysis.

ACKNOWLEDGMENTS

For computer time, this research used the resources of the Extreme Computing Research Center (ECRC) at KAUST. The authors thank the ECRC for supporting T. Malas. The authors gratefully acknowledge the support of the Erlangen Graduate School in Advanced Optical Technologies (SAOT) and the Cluster of Excellence "Engineering of Advanced Materials" at the University of Erlangen-Nuremberg, which are both funded by the German Research Foundation (DFG) in the framework of the German excellence initiative. The authors are also grateful for funding provided by the Energy Campus Nuremberg (EnCN, Project "Solarfabrik der Zukunft").